\documentclass[pra,twocolumn]{revtex4-2}
\usepackage{amsmath}
\usepackage{graphicx}

\newcommand{\bra}[1]{\left\langle #1\right|}
\newcommand{\ket}[1]{\left| #1\right\rangle}
\newcommand{\nangle}[1]{\left\langle #1 \right\rangle}

\renewcommand{\d}{{\rm d}}
\newcommand{\e}{{\rm e}}
\newcommand{\imai}{{\rm i}}

\usepackage{color}

\begin{document}
\title{Positivity of entropy production for the three-level maser}
\author{Alex Arash Sand Kalaee}
\author{Andreas Wacker}
\email{Andreas.Wacker@fysik.lu.se}
\affiliation{Mathematical Physics and NanoLund, Lund University, Sweden}
\date{\today, accepted manuscript, to appear in Physical Review A}

\begin{abstract}
Entropy production is a key concept of thermodynamics and allows one
to analyze the operation of engines. For the  Scovil-Schulz-DuBois heat engine, the
archetypal three-level thermal maser coupled to thermal baths, it was argued that
the common definition of heat flow may provide negative entropy production for certain parameters
[E. Boukobza and D. J. Tannor, Phys. Rev. Lett. 98, 240601 (2007)].
Here, we show that this can be cured, if 
corrections for detuning are properly applied to the energies used for the
bath transitions. This method can be used more generally for the thermodynamical analysis of
optical transitions where the modes of the light field are detuned from the transition energy.
\end{abstract}
\maketitle
\section{Introduction}

With the realization of masers and lasers quantum optics has proved fertile ground for thermodynamic research in open quantum systems.
An archetypal example is the Scovil-Schulz-DuBois heat engine based on a three-level maser driven by two heat baths of different temperatures \cite{ScovilPRL1959}. This system served as a model to develop a variety of approaches for the microscopic description of 
quantum systems in contact with thermal baths in interaction with classical \cite{LambPR1964,GevaPRE1994,GevaJCP1996,BoukobzaPRA2006b,ScullyPNAS2011} or quantized light fields \cite{BoukobzaPRA2006a,GhoshPNAS2018,NiedenzuQuantum2019}.

Of key relevance is the formulation of work and heat 
in the quantum realm.
Refs.~\cite{PuszCommMathPhys1978,AlickiJPA1979} defined work flow (power) and heat 
flow by partitioning the time derivative of the expectation value of the 
\textit{full} Hamiltonian, i.e. including the time-dependent interaction with 
classical degrees of freedom, such as a microwave field. Later, Boukobza and Tannor \cite{BoukobzaPRA2006b} proposed an alternative 
definition of power and heat flow by restricting to the \textit{bare} Hamiltonian, which describes the system itself
and lacks explicit time dependence. This is sometimes conceptually simpler and was, e.g., also used in Refs.~\cite{KirsanskasPRB2018,JaseemPRE2020}. The authors argued \cite{BoukobzaPRL2007}, that the bare
heat flows always provide a positive entropy production \cite{SpohnJMP1978} for the three level maser, 
while this was not the case for full heat flows in their treatment, which thus may violate the second law of thermodynamics. 
(We use full and bare in the sense that they
relate to the Hamiltonian from which the flows are derived.) 
The correct definition of heat and work is actually still an open issue, see, e.g., the discussion on page 339 of \cite{GelbwaserAdvAtomMolPhys2015}, where further references are given.

In this work we study the definitions for work and heat for the three-level maser coupled to a classical microwave field, where the bath couplings are treated by a Lindblad dissipator as outlined in Sec.~\ref{sec:system}.
Sec.~\ref{sec:retrace} focuses on the different definitions of heat and work, where we essentially follow Ref.~\cite{BoukobzaPRL2007}
showing a violation of entropy production for the full approach.
In Sec.~\ref{SecEffectiveEnergies} we present a reformulation of their expressions, which
allows for the correct identification of energies supplied by the baths. Using these
we recover a strictly positive entropy production for the full heat flows.

\section{The system}
\label{sec:system}
\begin{figure}
\centering
\includegraphics[width=\columnwidth]{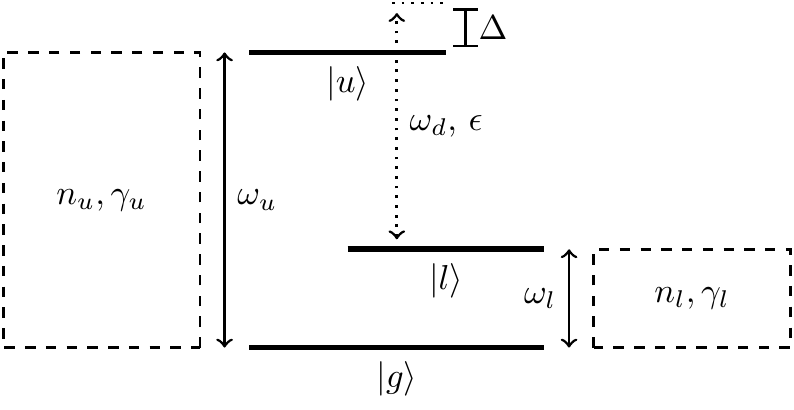}
\caption{Energy diagram of the  three-level maser subjected to a modulating field (dotted arrow), where the transitions $g\leftrightarrow u$ and  $g\leftrightarrow l$ (full arrows) are coupled to different reservoirs. 
A finite value of $\Delta = \omega_d - (\omega_u - \omega_l)$ reflects the detuning between the modulating field and the energy level difference.}
\label{fig:atom}
\end{figure}

We consider the three-level system of Scovil and Schulz-DuBois \cite{ScovilPRL1959} consisting of an upper ($u$) and lower ($l$) maser level and the
ground level ($g$), see Fig~\ref{fig:atom}. 
Throughout this article we set $\hbar=k_B=1$ in order to simplify the notation.
The full system Hamiltonian, $H = H_0 + V(t)$,
consists of the bare Hamiltonian $H_0 = \omega_u\sigma_{uu}+\omega_l\sigma_{ll}$
and a modulating external field
$V(t) = \epsilon (\e^{\imai\omega_d t}\sigma_{lu}+\e^{-\imai\omega_d t}\sigma_{ul})$,
where $\epsilon$ is the strength of the driving field, $\omega_d$ its modulating frequency, and we use the operators $\sigma_{ij} = \ket{i}\bra{j}$.
Without loss of generality the energy of the ground state $\ket{g}$ is set to zero.
The three-level system is connected to two bosonic reservoirs (baths), which are labeled by $\alpha$, where $\alpha\in\{u,l\}$.
The bath $\alpha$ 
couples to the transition $\ket{g}\leftrightarrow\ket{\alpha}$ with strength $\gamma_\alpha$, where 
an average number of excitations $n_\alpha$ is available in the bath.
The model and the analysis of its steady state behavior summarized below
follows recent work by \cite{BoukobzaPRA2006b,BoukobzaPRL2007,JaseemPRE2020}.

The time evolution of the system density operator $\rho$ is assumed to be Markovian
and governed by the Lindblad master equation \cite{LindbladCMP1976}
\begin{equation}
\dot\rho = -\imai[H(t),\rho]+\mathcal{L}_u[\rho]+\mathcal{L}_l[\rho]
\label{eq:system:lindblad}
\end{equation}
where the coupling to the baths are described by
$\mathcal{L}_\alpha[\rho] = \gamma_\alpha n_\alpha \mathcal{D}_{\sigma_{\alpha g}}[\rho]+\gamma_\alpha (n_\alpha+1)\mathcal{D}_{\sigma_{g\alpha}}[\rho]$
with the dissipator $\mathcal{D}_\sigma[\rho] = \sigma\rho\sigma^\dagger-
\frac{1}{2}\left\{\sigma^\dagger\sigma\rho+\rho\sigma^\dagger\sigma\right\}$.

To simplify the master equation we remove the time dependence of the Hamiltonian
by transforming the system to a rotating frame \cite{BoukobzaPRL2007,JaseemPRE2020}. For $X = \omega_l\sigma_{ll}+(\omega_l+\omega_d)\sigma_{uu}$, we define
$A^\textrm{rot} = U(t) A U^\dagger(t)$ according to the unitary operator
$U(t) = \e^{\imai Xt}$.
While the dissipative terms are unaffected by the choice of the rotating frame,
the unitary part of the quantum evolution is determined by the Hamiltonian
\begin{equation}
\widetilde{H} = H^\mathrm{rot}-X = -\Delta\sigma_{uu} + \epsilon(\sigma_{ul}+\sigma_{lu})
\end{equation}
with the detuning parameter $\Delta = \omega_d - (\omega_u-\omega_l)$.
Solving Eq.~\eqref{eq:system:lindblad} for the steady state
in the rotating frame (details are given in  App.~\ref{AppSystem})
yields  the net transition rate $R_{u\rightarrow l}$ from the upper to the lower level
\begin{equation}
  R_{u\rightarrow l}=  \frac{A(\gamma_u,\gamma_l,n_u,n_l,\epsilon)}{F(\gamma_u,\gamma_l,n_u,n_l,\epsilon,\Delta)}(n_u-n_l)
  \label{EqExpressionR}
\end{equation}
where the $A$ and $F$ are both positive  \cite{BoukobzaPRL2007}, see Eq.~\eqref{EqExpressionAF}. Thus
 $R_{u\rightarrow l}$ has the same sign as the difference $n_u-n_l$ between bath occupations which is driving the transitions.

\section{Work, heat, and entropy}
\label{sec:retrace}
Let the average energy in the system be $\nangle{E} = \mathrm{Tr}\{\rho H\}$.
The typical definitions of full power and full heat flows in the density matrix formalism are \cite{PuszCommMathPhys1978,AlickiJPA1979}
\begin{equation}
P = \dot{W} = \mathrm{Tr}\{\rho\dot{H}\},\quad\dot{Q} = \mathrm{Tr}\{\dot{\rho} H\}
\label{eq:typical}
\end{equation}
where we use the convention, that positive values of $P$ and $\dot{Q}$ correspond to an increase of energy in the system.

Alternatively, some authors apply an alternative definition of the
work and heat for systems coupled to a time-dependent external field
which is based on the bare Hamiltonian,
$\nangle{E_0} = \mathrm{Tr}\{\rho H_0\}$ \cite{BoukobzaPRA2006b,KirsanskasPRB2018}.
Based on the first law of thermodynamics the  bare flows are identified from
\begin{equation}
\dot{E_0} =-\imai\mathrm{Tr}\{\rho[H_0,V(t)]\}+\sum_{\alpha\in\{u,l\}}\mathrm{Tr}\{\mathcal{L}_\alpha[\rho]H_0\}
\label{EqHeatWorkBareHam}
\end{equation}
where the first (unitary) term is interpreted as the bare power $P_0$ and the second (dissipative)
term as the sum of bare heat flows $\dot{Q}_{0\alpha}$ from the respective baths to the system.
These terms can  be either evaluated in the original or the rotating frame due to the invariance of the trace under cyclic permutations of operators. (This is simpler compared to the first definition with full power and heat flow Eq.~\eqref{eq:typical}, where the transformations 
$[\dot{A}]^\textrm{rot}\neq \mathrm{d}A^\textrm{rot}/\mathrm{d} t$ for $A=H,\rho$ are more involved.)

From these definitions the steady state bare power and heat flow become (see App.~\ref{AppBare})
\begin{equation}
\begin{split}
P_0 &= -R_{u\rightarrow l}(\omega_u-\omega_l)\\
\dot{Q}_{0u} &= +R_{u\rightarrow l}\omega_u\\
\dot{Q}_{0l} &= -R_{u\rightarrow l}\omega_l
\label{eq:boukobza}
\end{split}
\end{equation}
We note that the bare power and heat flow correspond to the net transition rate
$R_{u\rightarrow l}$ multiplied with the respective bare transition energies from $H_0$.

The second law of thermodynamics requires a positive definite entropy production.
Spohn's entropy production function for the engine reads \cite{SpohnJMP1978}
\begin{equation}
\sigma = \frac{\partial S}{\partial t} - \frac{\dot{Q}_u}{T_u} - \frac{\dot{Q}_l}{T_l}
\label{eq:spohn}
\end{equation}
where $S=\mathrm{Tr}\{\rho\ln\rho\}=\mathrm{Tr}\{\rho^\textrm{rot}_\textrm{steady state}\ln\rho^\textrm{rot}_\textrm{steady state}\}$
is the von Neumann entropy \cite{VonNeumannBook1932} of the three-level system,
which is constant in steady state.
The temperatures of the baths are commonly related to the mean occupations as
\begin{equation}
T_\alpha = \frac{\omega_\alpha}{\log\left(1+\frac{1}{n_\alpha}\right)}
\label{eq:temperature}
\end{equation}
by using the appropriate Bose distribution function.

Using the  bare heat flows from Eq.~\eqref{eq:boukobza} Boukobza and Tannor found \cite{BoukobzaPRL2007}
\begin{equation}
\sigma_0 = R_{u\rightarrow l}\left[\log\left(1+\frac{1}{n_l}\right)-\log\left(1+\frac{1}{n_u}\right)\right] > 0
\label{eq:btentropy}
\end{equation}
which is positive definite as both factors have the same sign of $(n_u-n_l)$, see Eq.~\eqref{EqExpressionR}.
In contrast, using the full heat flows from Eq.~\eqref{eq:typical} with temperatures by Eq.~\eqref{eq:temperature},
Boukobza and Tannor\cite{BoukobzaPRL2007} detected negative entropy production for some operation points.
This suggested that the definition of work and heat based on the bare Hamiltonian \eqref{EqHeatWorkBareHam} 
should be preferred.

\section{Resolution by effective energies}\label{SecEffectiveEnergies}
Here, we rewrite the results for the full power and heat flows evaluated from Eq.~\eqref{eq:typical} in the form:
\begin{equation}
\begin{split}
P &= -R_{u\rightarrow l} \omega_d,\\
\quad\dot{Q}_u &= +R_{u\rightarrow l} \tilde{\omega}_u,\\
\quad\dot{Q}_l &= -R_{u\rightarrow l} \tilde{\omega}_l
\label{eq:typflux}
\end{split}
\end{equation}
with effective energies  (see App.~\ref{AppFull})
\begin{equation}
\begin{split}
\tilde{\omega}_{u}=\omega_{u}+\frac{\Delta\gamma_{u}(n_{u}+1)}{\gamma_u(n_u+1)+\gamma_l(n_l+1)}\\
\tilde{\omega}_{l}=\omega_{l}-\frac{\Delta\gamma_{l}(n_{l}+1)}{\gamma_u(n_u+1)+\gamma_l(n_l+1)}
\end{split}
\label{EqErenorm}
\end{equation}
Comparing Eq.~\eqref{eq:boukobza} with Eq.~\eqref{eq:typflux} we note that all flows are proportional to the transition rate $R_{u\to l}$, describing the round trip rate of the engine. However, there are different  
energy factors in each term.

For vanishing detuning, $\Delta = 0$, the respective energy factors in Eq.~\eqref{eq:boukobza} and Eq.~\eqref{eq:typflux} agree. Here, the heat fluxes from the baths are determined by the level energies 
$\omega_\alpha$ and the power transferred from the light field is given by the photon energy $\omega_d$, as expected.

However, for finite detuning, i.e. $\Delta =\omega_d-\omega_u+\omega_l\neq 0$, energy conservation does not allow for this structure, where the full and the bare approach provide different remedies: In the bare approach based on Eq.~\eqref{EqHeatWorkBareHam}, the power supplied from the ac field changes its energy factor $\omega_d\to \omega_u-\omega_l$, see Eq.~\eqref{eq:boukobza}. This appears not physical,
as a quantized ac field, should have energies in portions of $\hbar \omega_d$ and thus may result in an error of the order 
$\Delta R_{u\to l}$ in the power. In contrast, for the full approach,  the bare level energies are replaced by effective ones
$\omega_\alpha \to \tilde{\omega}_\alpha$,  see Eq.~\eqref{eq:typflux}, which satisfy $\omega_d=\tilde{\omega}_u-\tilde{\omega}_l$, so that energy conservation holds with the ac-frequency of the field.

Here we argue that the effective energies Eq.~\eqref{EqErenorm} should be taken seriously in the full approach
and thus be used in the definitions of the bath temperatures
\begin{equation}
\widetilde{T}_\alpha = \frac{\tilde{\omega}_\alpha}{\log\left(1+\frac{1}{n_\alpha}\right)}
\label{eq:temperatureNew}
\end{equation}
Then Eq.~\eqref{eq:spohn} provides
\begin{equation}
\begin{split}
  \sigma &= -\frac{\dot{Q}_u}{\widetilde{T}_u}-\frac{\dot{Q}_l}{\widetilde{T}_l}
  = R_{u\rightarrow l}\left[\frac{\tilde{\omega}_l}{\widetilde{T}_l}-\frac{\tilde{\omega}_u}{\widetilde{T}_u}\right]\\
 &=R_{u\rightarrow l}\left[\log\left(1+\frac{1}{n_l}\right)-\log\left(1+\frac{1}{n_u}\right)\right]
\end{split}
\end{equation}
which is identical with the entropy production function
Eq.~\eqref{eq:btentropy} from the bare approach and, most importantly, positive definite.

\begin{figure}
  \includegraphics[width=6cm]{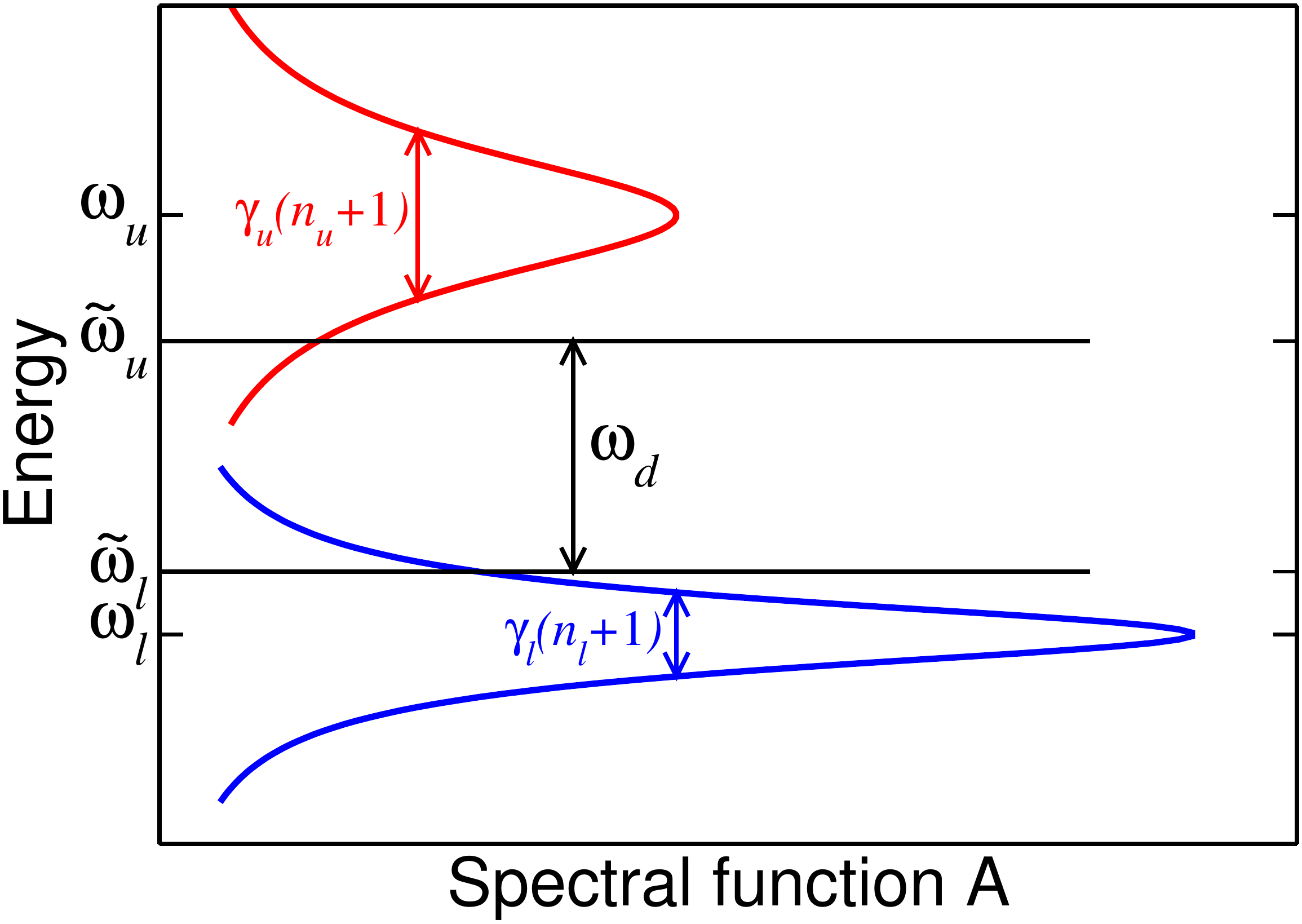}
  \caption{Sketch of the spectral functions \eqref{EqSpectralFunction} for the upper and lower levels, $u$ and $l$.
The black arrow shows the optical transition at frequency $\omega_d$ not matching the energy difference of the bare states.
The energies $\tilde{\omega}_u$ and $\tilde{\omega}_l$ are given by \eqref{EqErenorm}.}
  \label{FigSpectralFunctions}
\end{figure}

Now, we want to highlight the particular meaning of the energies $\tilde{\omega}_u$, $\tilde{\omega}_l$ from Eq.~\eqref{EqErenorm}.
Due to life-time broadening the energies of the
levels $u$ and $l$ are smeared out by Lorentzian spectral functions (here normalized to one)
\begin{equation}
A_\alpha(\omega)=\frac{1}{2\pi}\frac{\gamma_\alpha(1+n_\alpha)}{(\omega-\omega_\alpha)^2+\gamma_\alpha^2(1+n_\alpha)^2/4}
\label{EqSpectralFunction}
\end{equation}
with a full width at half maximum (FWHM) $\gamma_\alpha(1+n_\alpha)$ resulting from the decay of the states by relaxation to the ground level. This allows for energy-conserving
transitions between the levels $u$ and $l$ at the energy $\omega_d$
imposed by the ac field even if $\omega_d\neq\omega_u-\omega_l$, see Fig.~\ref{FigSpectralFunctions}.
Fermi's golden rule provides the transition rate from the initial level $u$ with the energy $\omega$
(similar but not necessarily equal to $\omega_u$).
\[
W_{u\to l}=2\pi\epsilon^2 A_l(\omega-\omega_d)
\]
Weighting with the density $A_u(\omega)$ of the initial state and multiplying with
the difference in occupation
$f_u(\omega)-f_l(\omega-\omega_d)$ of the levels (technically,  $f_\alpha$ is
the ratio between the imaginary part of the lesser Green's function and the spectral function\cite{HaugJauhoBook1996}),
we obtain the net transition rate
\begin{equation}
R_{u\to l}=2\pi\epsilon^2 \int d \omega\,  A_u(\omega)A_l(\omega-\omega_d)[f_u(\omega)-f_l(\omega-\omega_d)]\label{EqRbySpectral}
\end{equation}
Neglecting the energy dependence of $f_\alpha$ over the width of the spectral functions (which would be relevant to study dispersive/Bloch gain \cite{WackerNatPhys2007}), we set
$f_u(\omega)-f_l(\omega-\omega_d)\approx  \rho_{uu}-\rho_{ll}$. Then, some algebra, see Eq.~\eqref{EqConvolutionSpectral},
results in the expression \eqref{EqRate3Level}. This shows the equivalence of
this Green's function based treatment
with the density matrix calculations used above.

Eq.~\eqref{EqRbySpectral} shows that there is not a single definite energy involved for the upper and lower level, if broadening is taken
into account.
However, as the transitions occur with the weight $A_u(\omega)A_l(\omega-\omega_d)$,
we can identify the average energy for the upper level involved in transitions
\begin{equation}
  \langle \omega\rangle_u=\frac{\int d \omega\,  \omega A_u(\omega)A_l(\omega-\omega_d)}{\int d \omega  A_u(\omega)A_l(\omega-\omega_d)}
\end{equation}
and obtain after some algebra, see Eqs.~(\ref{EqConvolutionSpectral},\ref{EqConvolutionSpectralE}),
$\langle \omega\rangle_u=\tilde{\omega}_u$ with $\tilde{\omega}_u$ from Eq.~\eqref{EqErenorm}. 
The average energy for the lower level involved is then $\langle \omega\rangle_l=\langle \omega\rangle_u-\omega_d=\tilde{\omega}_l$.
Thus we find, that the effective levels from  Eq.~\eqref{EqErenorm} are the average energies involved in the optical transition,
if level broadening is taken into account. These are the average energies, which need to be added/removed from/to the respective bath
after a transition took place in order to restore the previous state. Therefore the bath properties at these energies is of most relevance which
justifies the definition of temperature via Eq.~\eqref{eq:temperatureNew}.

Energy exchange with the bath $\alpha$ at energies different from $\omega_\alpha$ requires that the
energies available in the baths cover a range of several $\gamma_\alpha$ around $\omega_\alpha$.
In the Green's function picture, this is the basis for assuming an energy-independent self-energy (i.e. a constant width in the spectral function).
For the Lindblad kinetics, the Markovian limit used requires a short bath correlation time $\tau_B\ll 1/\gamma_\alpha$ \cite{BreuerBook2006}
and consequently a spectral width of the bath well surpassing $\gamma_\alpha$.
This demonstrates again the consistency between the Green's function based treatment and the density matrix calculations.

Let us finally consider the Carnot efficiency of the engine.
Ref.~\cite{BoukobzaPRA2013} reported the occurrence of efficiencies
above $1-T_l/T_u$ for $\Delta >0$ in the semi-classical treatment of the ac
field. This is based on Eq.~(15)
of Ref.~\cite{BoukobzaPRA2013}, which (in our notation)
expresses the  efficiency as
\begin{equation}
  \eta=\frac{-P}{\dot{Q}_u}=\frac{\omega_d}{\tilde{\omega}_u}=1-\frac{\tilde{\omega}_l}{\tilde{\omega}_u}\, . \label{Eqeta}
\end{equation}
A positive power output $(-P)$ from the engine is based on $R_{u\to l}>0$
and thus requires $n_u>n_l$
by Eq.~\eqref{EqExpressionR}. Then our new definition of temperatures \eqref{eq:temperatureNew} provides
\[
n_u>n_l\Rightarrow \frac{\tilde{T}_u}{\tilde{\omega}_u}>\frac{\tilde{T}_l}{\tilde{\omega}_l}
\Leftrightarrow
\frac{\tilde{\omega_l}}{\tilde{\omega_u}}>\frac{\tilde{T_l}}{\tilde{T_u}}
\]
so that Eq.~\eqref{Eqeta} satisfies the Carnot efficiency  $\eta < 1-\tilde{T}_l/\tilde{T}_u$.

\section{Conclusion}
Both definitions of heat and work, applying either the full or the bare system Hamiltonian, provide identical (and positive definite) expressions for the entropy production for the common three-level maser driven by thermal baths. For the case of the full heat flow, it is crucial to carefully analyse the energies exchanged 
with the baths. These differ from the bare level energies if the ac field does not match the transition frequency. Disregarding this can provide violations of the second law as reported earlier \cite{BoukobzaPRL2007}.
While both the full and bare approach are thermodynamically consistent and provide identical expressions for entropy production, the full approach requires an elaborate description of the energies transferred to the bath, which rely on the steady state in our treatment. 
Furthermore, it is an open issue if such a description can be extended to transient behaviour, non-monochromatic fields, or non-cyclic operation \cite{BoukobzaPRA2013}.
On the other hand, the bare approach provides the transition frequency rather than the ac frequency in the work output, which provides a (typically small) error.

\section*{Acknowledgements}
We thank the Knut and Alice Wallenberg foundation and NanoLund for financial support.

\onecolumngrid
\begin{appendix}
\section{Detailed derivations for the steady state solution}
\label{AppSystem}
After transforming to the rotating frame, Eq.~\eqref{eq:system:lindblad} provides the
equations of motion for $\rho_{ij}=\langle i |{\rho}^\textrm{rot}|j\rangle$
\begin{eqnarray}
\frac{\d}{\d t} \rho_{gg}&=&\gamma_u (n_u+1)\rho_{uu}
+\gamma_l (n_l+1)\rho_{ll}-(n_u\gamma_u+n_l\gamma_l) \rho_{gg}\\
\frac{\d}{\d t} \rho_{uu}&=&\gamma_u n_u\rho_{gg}-\gamma_u (n_u+1)\rho_{uu}
+\imai\epsilon (\rho_{ul}-\rho_{ul}^*)\label{EqrhoLL}\\
\frac{\d}{\d t} \rho_{ll}&=&\gamma_l n_l\rho_{gg}-\gamma_l (n_l+1)\rho_{ll}
-\imai\epsilon (\rho_{ul}-\rho_{ul}^*)\label{EqrhoRR}\\
\frac{\d}{\d t} \rho_{ul}&=&\imai \Delta \rho_{ul}+\imai\epsilon(\rho_{uu}-\rho_{ll})-[\gamma_u(n_u+1)+\gamma_l(n_l+1)]\rho_{ul}/2
\label{EqrhoLR}
\end{eqnarray}
In the steady state (superscript $^\textrm{ss}$), Eq.~\eqref{EqrhoLR} provides
\begin{equation}
\rho^\textrm{ss}_{ul}=\frac{-\epsilon(\rho^\textrm{ss}_{uu}-\rho^\textrm{ss}_{ll})}{\Delta+\imai[\gamma_u(n_u+1)+\gamma_l(n_l+1)]/2}\label{EqrhoLRstat}
\end{equation}
Furthermore, we identify the net rate of transitions between  $u$ and $l$ due to the ac field:
\begin{equation}
R_{u\to l}=-\imai\epsilon (\rho^\textrm{ss}_{ul}-\rho^{\textrm{ss}*}_{ul})=C (\rho^\textrm{ss}_{uu}-\rho^\textrm{ss}_{ll})\quad \textrm{with }
C=\frac{\epsilon^2[\gamma_u(n_u+1)+\gamma_l(n_l+1)]}{[\gamma_u(n_u+1)+\gamma_l(n_l+1)]^2/4+\Delta^2}
\label{EqRate3Level}
\end{equation}
Using $\rho_{gg}=1- \rho_{uu}- \rho_{ll}$, 
Eqs.~(\ref{EqrhoLL},\ref{EqrhoRR},\ref{EqRate3Level}) provide
the system of equations
\[\begin{split}
\gamma_u n_u=& [\gamma_u(2n_u+1)+C]\rho^\textrm{ss}_{uu}+(\gamma_un_u-C)\rho^\textrm{ss}_{ll}\\
\gamma_l n_l=& (\gamma_l n_l-C)\rho^\textrm{ss}_{uu}+[\gamma_l(2n_l+1)+C]\rho^\textrm{ss}_{ll}
\end{split}\]
with the solution
\[\begin{split}
\rho^\textrm{ss}_{uu}&=\frac{\gamma_u\gamma_ln_u(n_l+1)+C(\gamma_un_u+\gamma_l n_l)}{[\gamma_l(2n_l+1)+C] [\gamma_u(2n_u+1)+C]- (\gamma_l n_l-C)(\gamma_u n_u-C)}\\
\rho^\textrm{ss}_{ll}&=\frac{\gamma_u\gamma_ln_l(n_u+1)+C(\gamma_un_u+\gamma_l n_l)}{[\gamma_l(2n_l+1)+C][\gamma_u(2n_u+1)+C]- (\gamma_l n_l-C)(\gamma_u n_u-C)}
\end{split}\]
so that
\[
\rho^\textrm{ss}_{uu}-\rho^\textrm{ss}_{ll}=\frac{\gamma_l\gamma_u (n_u-n_l)}{\gamma_u\gamma_l(3n_u n_l+2n_u+2n_l +1)+C[\gamma_u(3n_u+1)+\gamma_l(3n_l+1)]}
\]
is proportional to the occupation differences of the baths. Inserting into Eq.~\eqref{EqRate3Level}, we obtain
Eq.~\eqref{EqExpressionR} from the main article, where
\begin{equation}
\begin{split}
A=&\frac{\gamma_l\gamma_u}{4}[\gamma_u(n_u+1)+\gamma_l(n_l+1)]\epsilon^2\\ 
F=&\frac{\gamma_u(n_u+1)+\gamma_l(n_l+1)}{2}\frac{\gamma_u(3n_u+1)+\gamma_l(3n_l+1)}{2}\epsilon^2\\
&+\frac{\gamma_l\gamma_u}{4}(3n_u n_l+2n_u+2n_l +1)
\left\{\frac{\left[\gamma_u(n_u+1)+\gamma_l(n_l+1)\right]^2}{4}+\Delta^2\right\}
\label{EqExpressionAF}
\end{split}\end{equation}
are quadratic polynomials in $\epsilon$. Thus the rate
$R_{u\to l}\propto \epsilon^2$ for small coupling $\epsilon$ to the ac field,
while it saturates for $\epsilon^2\gtrsim \gamma_u^2 (n_u+1)^2+\gamma_l^2 (n_l+1)^2+4\Delta^2$.
$A$ and $F$ are identical with the expressions in Eq.~(13) of \cite{BoukobzaPRL2007}, where
$\gamma_{0\alpha}=\gamma_{\alpha}/2$ is used.

\section{Heat and work with bare Hamiltonian}
\label{AppBare}
The definition of heat flow from the bare Hamiltonian \eqref{EqHeatWorkBareHam} provides the
bare heat entering from bath $u$ (note that the energy of the ground level is zero)
\begin{equation}
\dot{Q}_{0u}= \omega_u\langle u|{\cal L}_u({\rho})|u\rangle
=\omega_u\left[\gamma_u n_u\rho_{gg}-\gamma_u(n_u+1)\rho_{uu}\right]\label{EqHeat1bareL}
\end{equation}
Note that the diagonal elements of ${\rho}(t)$ 
are identical in the original and rotating frame.
Thus, in the steady state, Eq.~\eqref{EqrhoLL} provides
$\dot{Q}_{0u}^\textrm{ss}=-\imai \omega_u\epsilon (\rho^\textrm{ss}_{ul}-\rho^\textrm{ss}_{lu})
=\omega_uR_{u\to l}$
and similarly we get $\dot{Q}_{0l}^\textrm{ss}=-\omega_l R_{u\to l}$.
Finally, the bare work  \eqref{EqHeatWorkBareHam} done by the field on our systems is 
\[
P_0=
\imai \textrm{Tr}\{{\rho}[{V}(t),{H}_0]\}=
\imai \textrm{Tr}\{{\rho}^\mathrm{rot}[{V}^\mathrm{rot},{H}_0^\mathrm{rot}]\}=
\imai\epsilon(\omega_u-\omega_l)(\rho_{ul}-\rho_{ul}^*)
\]
which in the steady state provides $P_0^\textrm{ss}=-(\omega_u-\omega_l)R_{u\to l}$
so that $\dot{Q}_{0u}^\textrm{ss}+\dot{Q}_{0l}^\textrm{ss}+P^\textrm{ss}_0=0$, as required by energy conservation.
These are the terms provided in Eq.~\eqref{eq:boukobza} without the superscript $^\textrm{ss}$.

\section{Heat and work with full Hamiltonian}
\label{AppFull}
With the definition \eqref{eq:typical}, we obtain the power transferred to the system
\[
P(t)=\imai\epsilon\omega_d \textrm{Tr}\left\{{\rho} \left(|l\rangle \langle u|\e^{\imai\omega_d t}-
|u\rangle \langle l|\e^{-\imai\omega_d t}\right)\right\}=
\imai\epsilon \omega_d (\rho_{ul} -\rho_{lu})
\stackrel{\textrm {in ss}}{=}-\omega_dR_{u\to l}
\]
which corresponds to the net rate of absorbed photons $(-R_{u\to l})$ times the photon energy $\omega_d$.  (Note, that  we defined $\rho_{ul}$ in the rotating frame, see App.~\ref{AppSystem}, so that $\rho_{ul}=\langle u |{\rho}^\textrm{rot}|l\rangle=\langle u |{\rho}|l\rangle\e^{\imai\omega_d t}$.)

For the heat flow, the unitary evolution of ${\rho}(t)$ due to the Hamiltonian does not contribute, as
$\textrm{Tr}\left\{[{\rho},H]H\right\}=\textrm{Tr}\left\{{\rho}[H,H]\right\}=0$,
  where we used the invariance of the trace under cyclic permutations.
  Thus we can restrict to the non-unitarian part here. Then the part with $H_0$ provides the heat current $\dot{Q}_{0u}$ from Eq.~\eqref{EqHeat1bareL}. We have to add the part with ${V}(t)$ and find 
\[
\dot{Q}_{u}=\dot{Q}_{0u}
+\epsilon
\langle u|{\cal L}_u[{\rho}]|l\rangle\e^{\imai\omega_d t}
+\langle l|{\cal L}_u[{\rho}]|u\rangle\e^{-\imai\omega_d t}
=\dot{Q}_{0u}
-\epsilon\frac{\gamma_u(n_u+1)}{2} (\rho_{ul}+ \rho_{lu})
\]
Using Eqs.~(\ref{EqrhoLRstat},\ref{EqRate3Level}) we get in the steady state
\begin{eqnarray}
\dot{Q}^\textrm{ss}_{u}&=&\dot{Q}_{0u}^\textrm{ss}-
\frac{\gamma_u(n_u+1)}{2} 
\frac{\Re\left\{\rho^{ss}_{ul}\right\}}{\Im\left\{\rho^{ss}_{ul}\right\}} R_{u\to l}
= R_{u\to l} \tilde{\omega}_u \label{EqQ1}\\
\dot{Q}_{l}^\textrm{ss}&=&\dot{Q}_{0l}^\textrm{ss}-
\frac{\gamma_l(n_l+1)}{2} 
\frac{\Re\left\{\rho^{ss}_{ul}\right\}}{\Im\left\{\rho^{ss}_{ul}\right\}} R_{u\to l}
= -R_{u\to l} \tilde{\omega}_l \label{EqQ2}\\
\textrm{with } \tilde{\omega}_u&=&\omega_u +\frac{\Delta\gamma_u(n_u+1)}{\gamma_u(n_u+1)+\gamma_l(n_l+1)},\quad \tilde{\omega}_l=\omega_l -\frac{\Delta\gamma_l(n_l+1)}{\gamma_u(n_u+1)+\gamma_l(n_l+1)}\label{EqQEnergies}
\end{eqnarray}
where $\Re\{z\}$ and $\Im\{z\}$ denote, respectively, the real and imaginary part of a complex value $z$.
The full power and heat flow satisfy energy conservation $P^\textrm{ss}+\dot{Q}_{u}^\textrm{ss}+\dot{Q}_{l}^\textrm{ss}=0$ and 
provide Eqs.~(\ref{eq:typflux},\ref{EqErenorm}), where we omitted the superscript $^\textrm{ss}$.

\section{Convolution of Lorentzians}
We consider the function
\[
P(\omega,\Delta)=\frac{1}{2\pi}\frac{2\gamma_u}{\omega^2+\gamma_u^2}\frac{2\gamma_l}{(\omega-\Delta)^2+\gamma_l^2}
\]
which is the product of two spectral functions with FWHM $2\gamma_\alpha$.
Then we find with the residue theorem
\begin{equation}\begin{split}
\int\d \omega\, P(\omega,\Delta)=& \imai\left[\frac{2\gamma_u}{2\imai\gamma_u}
\frac{2\gamma_l}{(\imai\gamma_u-\Delta)^2+\gamma_l^2}+\frac{2\gamma_u}{(\Delta+\imai\gamma_l)^2+\gamma_u^2}
\frac{2\gamma_l}{2\imai\gamma_l}\right]\\
=&  \frac{2\gamma_l\left[(\Delta+\imai\gamma_l)^2+\gamma_u^2\right]+2\gamma_u\left[(\Delta-\imai\gamma_u)^2+\gamma_l^2\right]}
{\left[(\Delta-\imai\gamma_u)^2+\gamma_l^2\right]\left[(\Delta+\imai\gamma_l)^2+\gamma_u^2\right]}
\\
=& \frac{2(\gamma_u+\gamma_l)\left[\Delta^2-2\imai\Delta(\gamma_u-\gamma_l)-(\gamma_u-\gamma_l)^2\right]}
{\left[\Delta^2+(\gamma_u+\gamma_l)^2\right]\left[\Delta^2-2\imai\Delta(\gamma_u-\gamma_l)-(\gamma_u-\gamma_l)^2\right]}
= \frac{2(\gamma_u+\gamma_l)}{\Delta^2+(\gamma_u+\gamma_l)^2}
\label{EqConvolutionSpectral}
\end{split}\end{equation}
where the third identity is verified by comparing the results of the products in numerator and denominator, respectively.
The main result is that we obtain a Lorentzian with the sum of the individual widths.
Similarly we find
\begin{equation}\begin{split}
\int\d \omega\,\omega P(\omega,\Delta)=& \imai\left[\frac{2\gamma_u}{2\imai\gamma_u}
\frac{\imai\gamma_u 2\gamma_l}{(\imai\gamma_u-\Delta)^2+\gamma_l^2}+\frac{2\gamma_u(\Delta+\imai\gamma_l)}{(\Delta+\imai\gamma_l)^2+\gamma_u^2}
\frac{2\gamma_l}{2\imai\gamma_l}\right]\\
=&  \frac{2\imai\gamma_u\gamma_l\left[(\Delta+\imai\gamma_l)^2+\gamma_u^2\right]+2\gamma_u(\Delta+\imai\gamma_l)\left[(\Delta-\imai\gamma_u)^2+\gamma_l^2\right]}
{\left[(\Delta-\imai\gamma_u)^2+\gamma_l^2\right]\left[(\Delta+\imai\gamma_l)^2+\gamma_u^2\right]}
\\
=& \frac{2\gamma_u\Delta\left[\Delta^2-2\imai\Delta(\gamma_u-\gamma_l)-(\gamma_u-\gamma_l)^2\right]}
{\left[\Delta^2+(\gamma_u+\gamma_l)^2\right]\left[\Delta^2-2\imai\Delta(\gamma_u-\gamma_l)-(\gamma_u-\gamma_l)^2\right]}
= \frac{2\gamma_u\Delta}{\Delta^2+(\gamma_u+\gamma_l)^2}
\label{EqConvolutionSpectralE}
\end{split}\end{equation}

\end{appendix}

\twocolumngrid
%

\end{document}